\documentclass[aps,pre,reprint,twocolumn,showpacs,superscriptaddress,longbibliography]{revtex4-2}
\usepackage{graphicx}% Include figure files
\usepackage{dcolumn}% Align table columns on decimal point
\usepackage{bm}% bold math
\usepackage{amsmath}
\usepackage{diagbox}

\usepackage{hyperref}
\hypersetup{
    colorlinks=true,
    urlcolor=blue
}

\usepackage{color}
\usepackage{xcolor}

\usepackage{multirow}

\usepackage{placeins}
\usepackage[english]{babel}
\begin{document}

\title{From Radiation Dose to Cellular Dynamics:\\ A Discrete Model for Simulating Cancer Therapy}

\author{Mirko Bagnarol}
\email[]{mirko.bagnarol@mail.huji.ac.il}
%\homepage[]{Your web page}
%\thanks{}
%\altaffiliation{}
\affiliation{The Racah Institute of Physics, The Hebrew University of Jerusalem, Jerusalem 9190401, Israel}

\author{Gianluca Lattanzi}
\email[]{gianluca.lattanzi@unitn.it}
%\homepage[]{Your web page}
%\thanks{}
%\altaffiliation{}
\affiliation{Department of Physics, University of Trento and INFN-TIFPA, Trento Institute for Fundamental Physics and Applications, Via Sommarive 14, Povo (TN) 38123, Italy}

\author{Jan {\AA}str{\"o}m}
\email[]{jan.astrom@csc.fi}
%\homepage[]{Your web page}
%\thanks{}
%\altaffiliation{}
\affiliation{CSC Scientific Computing Ltd, K{\"a}gelstranden 14, 02150 Esbo, Finland}

\author{Mikko Karttunen}
\email[]{mikko.karttunen1@uef.fi}
%\homepage[]{https://www.softsimu.net/mikko/}
%\thanks{}
%\altaffiliation{}
\affiliation{European Laboratory for Learning and Intelligent Systems (ELLIS) Institute Finland, Maarintie 8, 02150 Espoo, Finland}
\affiliation{Department of Technical Physics, University of Eastern Finland, P.O. Box 1627, FI-70211 Kuopio, Finland}
\affiliation{Department of Physics and Astronomy,
  Western University, 1151 Richmond Street, London,
  Ontario, Canada  N6A\,3K7}
\affiliation{Department of Chemistry,  Western University, 1151 Richmond Street,
London, Ontario, Canada N6A\,5B7}

\date{\today}

\begin{abstract}
Radiation therapy is one of the most common cancer treatments, and dose optimization and targeting of radiation are crucial since both cancerous and healthy cells are affected. Different mathematical and computational approaches have been developed for this task. The most common mathematical approach, dating back to the late 1970's, is the linear-quadratic (LQ) model for the survival probability given the radiation dose. Most simulation models consider tissue as a continuum rather than consisting of discrete cells. 
%
%\textcolor{red}{
While reasonable for large-scale models (e.g., human organs), continuum approaches necessarily neglect cellular-scale effects, which may play a role in growth, morphology, and metastasis of tumors.
%} 
%
Here, we propose a method for modeling the effect of radiation on cells based on the mechanobiological \textsc{CellSim3D} simulation model for growth, division, and proliferation of cells. To model the effect of a radiation beam, we incorporate a Monte Carlo procedure into \textsc{CellSim3D} with the LQ model by introducing a survival probability at each beam delivery. Effective removal of dead cells by phagocytosis was also implemented. Systems with two types of cells were simulated: stiff slowly proliferating healthy cells and soft rapidly proliferating cancer cells. For model verification, the results were compared  to prostate cancer (PC-3 cell line) data for different doses and we found good agreement. In addition, we simulated proliferating systems and analyzed the probability density of the contact forces. We determined the state of the system with respect to the jamming transition and found very good agreement with experiments.
\end{abstract}

\maketitle

\section{Introduction}

According to the US Centers for Disease Control and Prevention (CDC), cancer was the second most
common cause of death responsible for 21.9\% and 20.7\% of deaths of males and females, respectively, in 2017~\cite{cdc2017female, cdc2017male}. It has been estimated that as many as 70\% of cancer patients receive radiotherapy either as the main treatment or in combination with others~\cite{Blyuss2015}. Radiation treatment is particularly effective in breast and prostate cancers, and according to Bryant \textit{et al.}~\cite{Bryant2017}, 40\% of breast and 23\% of prostate cancer survivors received
radiation treatment.

Radiotherapy can be described as a treatment with the aim to dysfunctionalize cancer cell DNA by damaging it by irradiation of targeted high-energy beam of, e.g., x-rays, protons or electrons. Radiation is not selective and also damages healthy cells and their DNA, and thus may, in addition to immediate side effects, have side effects that take long times to be manifest. Due to the above reasons, optimization of spatial accuracy, dose and treatment frequency are critical for success. Reviews of the perspectives and history of radiotherapy are provided, for example, by Bernier \textit{et al.}~\cite{Bernier2004}, Connell and Hellman~\cite{Connell2009}, and Streffer and Herrmann~\cite{Streffer2012}. Different approaches to modeling radiotherapy are provided, e.g., Jones and Dale~\cite{Jones2019-tm}.

There are different approaches to treatment planning algorithms. They are typically divided into correction-based and model-based, the latter being generally more accurate~\cite{Chen2014j,Woon2016}. Difficulties include, for example, inhomogeneous tissue, fractonation of dose, over-expression of self-repair mechanisms in cancer cells and cell migration. One further issue is that most models treat the tissue as a continuum rather than consisting of cells of discrete sizes and varying mechanical properties. It has recently been shown that the local \textit{mechanical stress landscape} has a strong influence on cell migration, and that targeting mechanical properties may provide a new pathway to cancer treatment~\cite{Tambe2011a}.

Some of the current modeling approaches include agent-based ones~\cite{Jalalimanesh2017} which may offer a new and adaptable way for modeling; although not specific to radiotherapy, many of the current models for cancer are discussed in the editorial by Enderling and Rejniak~\cite{Enderling2013} and other articles in that same special issue. Other reviews discussing radiotherapy related matters are provided, e.g., by Marcu and Harriss-Phillips~\cite{Marcu2012}, and D'Andrea \textit{et al.}~\cite{DAndrea2016} As the reviews show, discreteness of cells is very rarely accounted for in radiotherapy modeling. One of the models in that direction is the one by Bobadilla \textit{et al.}~\cite{Bobadilla2017} where  cells are differentiated according to whether they have received radiation therapy or not. Conceptually, we combine different levels of modeling using a mechanobiological cell model together with the continuum LQ model. This is in the same sprit as the work of Liu \textit{et al.}~\cite{Liu2020-hg}, who combined radiation transport using the Geant4 framework, and a cell simulation platform called CompuCell3D.

The aim of this paper is to develop a radiotherapy model based on discrete cells with heterogeneous properties, each capable of responding  to changes in its local environment. To this end, we used the \textsc{CellSim3D} model~\cite{Madhikar2018} as the basis, and integrated features of an external radiation beam by combining it with a Monte Carlo process simulation and the LQ approach. Using  \textsc{CellSim3D} as the basis also provides an easily expandable tool beyond LQ. It is important to note that the current implementation assumes a spatially uniform or photon-based radiation field, which aligns with standard clinical photon therapies. Extending this framework to account for more complex radiation types, such as high-energy protons or heavier ions, remains an open and technically challenging problem. Preliminary studies in this direction exist~\cite{Cogno2024,Garcia2024}, particularly for high-energy protons where the biological effects tend to resemble those of photons under certain conditions. Nonetheless, accurately modeling the stochastic and spatially heterogeneous energy deposition characteristic of particle therapies would require significant advancements in both physical modeling and biological response integration, and represents a promising avenue for future work.
%
%{\color{red}
\begin{table*}[t]
    \centering
    \begin{tabular}{|c|c|c|c|c|c|c|}
     \toprule
    \textbf{Simulation} & \textbf{Total \# of steps} ($t_{\text{end}}$) & \textbf{Write frequency} ($\Delta t_{\text{write}}$) & \textbf{Box side} & \textbf{Dose} [Gy] & $\alpha$ [Gy$^{-1}$] & $\beta$ [Gy$^{-2}$]  \\
    \hline \hline
    \multirow{15}{*}{Prostate} & \multirow{15}{*}{7.25$\cdot10^{5}$} & \multirow{15}{*}{2.5$\cdot10^{4}$}  & \multirow{15}{*}{330} & 3  & \multirow{5}{*}{0.064} & \multirow{5}{*}{0.0167} \\
    \cline{5-5}
    & & & & 6 &  &  \\
    \cline{5-5}
    & & & & 9 &  &  \\
    \cline{5-5}
    & & & & 15 &  &  \\
    \cline{5-5}
    & & & & 20 &  &  \\
    \cline{5-7}
    & & & & 3 & \multirow{5}{*}{0.241} & \multirow{5}{*}{0.067} \\
    \cline{5-5}
    & & & & 6 &  &  \\
    \cline{5-5}
    & & & & 9 &  &  \\
    \cline{5-5}
    & & & & 15 &  &  \\
    \cline{5-5}
    & & & & 20 &  &  \\
    \cline{5-7}
    & & & & 3 & \multirow{5}{*}{0.487} & \multirow{5}{*}{0.055} \\
    \cline{5-5}
    & & & & 6 &  &  \\
    \cline{5-5}
    & & & & 9 &  &  \\
    \cline{5-5}
    & & & & 15 &  &  \\
    \cline{5-5}
    & & & & 20 &  &  \\
    \hline
    \multirow{8}{*}{Force distribution} & \multirow{8}{*}{5.5$\cdot10^{5}$}  & \multirow{8}{*}{2$\cdot10^{4}$} & \multirow{4}{*}{47} & 0 & \multirow{8}{*}{1} & \multirow{8}{*}{1} \\
    \cline{5-5}
    & & & & 0.2333 &  &  \\
    \cline{5-5}
    & & & & 0.4712 &  &  \\
    \cline{5-5}
    & & & & 0.7792 &  &  \\
    \cline{4-5}
    & & & \multirow{4}{*}{500} & 0 &  &  \\
    \cline{5-5}
    & & & & 0.2333 &  &  \\
    \cline{5-5}
    & & & & 0.4712 &  &  \\
    \cline{5-5}
    & & & & 0.7792 &  &  \\
    \hline \hline
    \end{tabular}
    \caption{The main simulation parameters: 
    $t_{\text{end}}$ defines the total number of time steps and $\Delta t_{\text{write}}$ is the frequency for saving data. The box side is given in simulation units, with one unit $\approx 10\ \mu$m (set by $V_{\text{div}}$ for HeLa cells). Thus, sides of 47, 330, and 500 correspond to $\approx 0.47$ mm, $3.3$ mm, and $5$ mm, respectively. The dose $D$ and the parameters $\alpha$ and $\beta$ describe the action of the beam and the radiation resistance of the tissue according to Eq.~\eqref{eq:surv}. Note that a dose of zero means that the beam is absent. The final number of cells in each calculation varied between $2.5\cdot 10^4$ to $5\cdot 10^4$.}
\label{tab:params} 
\end{table*}
%}

\section{Methods}
\subsection{CellSim3D and inclusion of radiation beam}

%{\color{red}
The \textsc{CellSim3D}~\cite{Madhikar2018} model and software was used as the basis for our simulations. \textsc{CellSim3D} is an open-source, GPU-accelerated framework for mechanobiological simulations of cell populations~\cite{cellsim3d-web}.  It represents each cell as a deformable elastic shell based on a C$_{180}$ fullerene geometry, where nodes on the surface interact through bonded and non-bonded forces. The total force acting on a node is given by the equation 
\begin{equation}
m\ddot{\mathbf{r}} = F_\text{bond} + F_\theta + F_\text{pressure} + F_\text{rep} + F_\text{adh} + F_\text{friction},
\end{equation}
where the terms correspond to internal elasticity, curvature preservation, osmotic pressure, short-range repulsion, adhesion between neighboring cells, and friction with other cells and the medium~\cite{Madhikar2018, Mazarei2022-vu}. When a cell’s volume exceeds a threshold $V_\text{div}$, it divides into two daughter cells according to a prescribed division plane, modeling mitosis.  

Although detailed derivations and parameter mappings to experimental systems are provided in previous work~\cite{Madhikar2018, kart2, Madhikar2020b,Mazarei2022-vu}, here we emphasize that \textsc{CellSim3D} routinely handles aggregates of $10^5$ cells in three dimensions on a single desktop GPU. 
The model includes cell-cell adhesion, intercellular and cell-medium friction, migration, growth, and division, with easily adjustable elastic and mechanical properties. 
This combination makes it well suited for studying the collective mechanics of growing cell populations within physically realistic constraints.
%}

Here, we have amended the original model by adding 1) the effect of a generic therapeutic radiation beam, 2) effective phagocytosis, and 3) the capability of simulating more than one cell type (here, healthy and cancer cells were used).
The effect of the  radiation beam is implemented as a random process: it acts by picking a random number $\xi _n$  drawn for each cell. This number is then compared with the \textit{death probability} $D_n$. If $\xi _n < D_n $ the $n^\mathrm{th}$ cells dies and cannot reproduce anymore. Other properties are left unchanged. This is essentially a simple Monte Carlo step.

\subsection{Linear-Quadratic (LQ) model}

The death probability ($D_n$) is taken from the Linear-Quadratic (LQ) model that was first introduced in the 1970's~\cite{Kellerer1972-lf,Chadwick1973-qe,Kellere1978-ns}. The LQ model remains the most frequently used one~\cite{Dale1985,ORourke2008,Brenner2008,Santiago2016,Leeuwen2018,Jones2019-tm}, and it is typically considered to work for doses in the range of 2-15\,Gy.
It assumes that the \textit{survival probability} of a cell can be expressed as
\begin{equation}
    S(D) = e^{-(\alpha D + \beta D^2)},
    \label{eq:surv}
\end{equation}
where $D$ is the dose delivered by the beam, while $\alpha$ ($[\alpha ] = \mathrm{Gy}^{-1}$) and $\beta$  ($[\beta ] = \mathrm{Gy}^{-2}$), called \textit{survival parameters}, describe the radiosensitivity of the tissue. The larger these parameters, the more sensitive the cells are to radiation~\cite{Mcmahon2018}. The parameter $\alpha$ is considered to model a single-track process meaning that the cellular DNA is damaged in a single radiation event. The parameter $\beta$ describes a situation of a two-track process in which the DNA is damaged by two separate radiation hits. Thus, the ratio $\alpha/\beta$ is an important characteristic of the model, providing the relative importance of single and double events.
%
%\textcolor{red}{
Consequently, for the $n$-th cell, $D_n = 1 - S_n(D)$, where the label $n$ in $S_n$ refers to potentially different values of $\alpha$ and $\beta$ for different cells. If the survival parameters are the same for all cells, so is the death probability $D_n$.
%}

The parameters $\alpha$ and $\beta$ are experimentally accessible (albeit there can be important differences between \textit{in vitro} measurements and \textit{in vivo} outcomes) and they both depend on the nature of the beam and the target tissue. The rationale for Eq.~\ref{eq:surv} is that it is a mechanistic model describing radiation damage to the DNA caused by two different types of processes. More complex approximations for $S(D)$, including those based on the above LQ model, Lea-Catcheside factors, or extensions of the LQ model to low and high dosages are not considered here, but they can be easily implemented in the code. For generalizations of the LQ model, see, e.g., Refs.~\cite{Brenner2008,ORourke2008,Leeuwen2018,Cordoni2021, Cordoni2022, Missiaggia2024}.

\subsection{The action of the radiation beam}

The action of the beam during a time step is instantaneous. Once the program determines the set of dying cells, their status is switched from \textit{alive} to \textit{dead} during the same time step. In real treatments, a single dose delivery usually lasts from 10 to 30 minutes, and thus the instantaneous action may seem as a harsh approximation. This is, however, not the case due to time scales. In particular, the time unit in \textsc{CellSim3D} is determined by cellular growth~\cite{Madhikar2018}. The program saves the state of
the system in the output file once every $t_{\text{write}}$ steps, where $t_{\text{write}}$, namely the \textit{output frame frequency}, is a number chosen by the user. Therefore, if the time interval between two successive frames reflects a period longer than 30 minutes, the beam can be considered as instantaneous.

In our simulations, the life cycle of a cancerous cell defined from the newborn state and ending with division, lasts less than 1,500 integration steps. In real time scales, the cell cycle of a human cancerous cell has a duration of about 24-72 hours, mitosis taking about 1 hour, setting one single frame to be 
in that range. Therefore, since one frame stands for a period longer than any dose delivery, cell death can be thought as instantaneous.

%{\color{red}
Moreover, the natural length scale is set by the division volume $V_{\text{div}}$, with $V_{\text{div}} = 2.25$ corresponding to a HeLa cell of volume $\sim 2{,}40\,\mu$m$^3$~\cite{Zhao2008}. This mapping implies that one unit of simulation length is about $10\,\mu$m. 
In the simulations with the smallest box (side 47 units, $\approx 0.47$\,mm), the final number of cells was below $2.5\times 10^4$, resulting in a dense, confined configuration. In contrast, in the larger boxes (sides 330–500 units, i.e. 3.3–5.0\,mm) the aggregates formed nearly spherical clusters of about $5\times 10^4$ cells that did not reach the boundaries. At a typical cellular packing fraction of $\sim 0.6$, such a cluster occupies a volume of $\sim 0.05$\,mm$^3$, corresponding to an effective diameter of about 0.45\,mm.
Since clinical photon and proton beams exhibit spatial dose variations only on mm to cm  scales~\cite{Khan2020}, well above the size of our simulated tumors, it is a very good approximation to assume a spatially uniform radiation beam (i.e. constant dose $D$) throughout the simulation volume.
%}

In reality, dead cells undergo phagocytosis~\cite{phago1}. We modeled this process as follows: when phagocytosis is activated, the program progressively lowers the internal pressure of the dead cells, until the threshold defined by the user is reached. The rationale is that by decreasing the volume  of the dead cells they become very soft and flexible, and free up space for the live cells to proliferate. The effect of this is essentially the same as the full removal of the dead cells but without the computational complexity required by restoring the topology of the system and neighbor tables. Phagocytosis is, in general, much faster than cellular reproduction~\cite{Gregory2011, Elliott2016} and hence the shrinking rate for the dead cells is set to be ten times the growth rate of alive cells.
%
%\textcolor{red}{
%We chose a shrinkage rate ten times higher than the cell growth rate.
This ensures that phagocytosis (removal of dead cell volume) occurs much faster than cell division, preserving the correct temporal order of events. The factor can be adjusted if needed, but maintaining a clear separation of timescales is crucial:
%}
%
while the unit of time for the simulation is determined by the cell cycle~\cite{kart2}, it is important to keep the events that occur within a single simulation in the right order.

In the current approach, two different cell types are present simultaneously. The first cell type represents healthy and the second one cancerous cells; we note that the type of cells is not limited to two, but can be varied as necessary. The two types of cells are identical in all parameters but Young's modulus, which describes the cell stiffness; cancer cells are typically considered to be softer than healthy cells~\cite{Cross2007, Plodinec2012, Alibert2017}. Cells with a higher Young's modulus are less prone to deformation and therefore grow slowly when compared to softer cells. Slower growth results in slower division, which mirrors the well-known fact that cancer cells proliferate faster than  healthy ones~\cite{Fabrikant1969, hanahan2011}. In prior studies this approach has been successfully  applied to study tumor growth in epithelial tissues~\cite{Madhikar2020a}.

\begin{figure}
    \centering
    \includegraphics[width=\columnwidth]{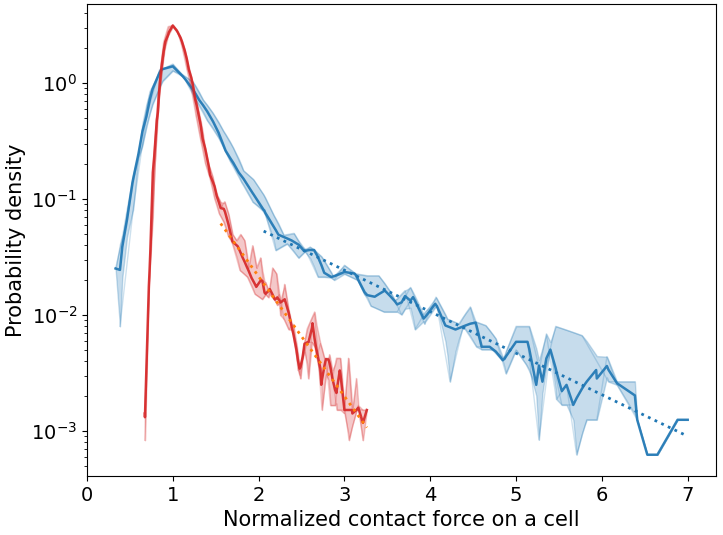}
    \caption{%
%    \textcolor{red}{
    Distribution of the normalized contact force of dead cells in the packed (red) and open (blue) spatial configurations. The shaded area contains the three curves with three different fractions of dead cells in the final frame of the simulation, being $D_1 \approx 20$\%, $D_2 \approx 40$\% and $D_3 \approx 60$\%, while the solid line is the median of the three. The broken line is an exponential fit using Eq. \eqref{eq:fitExp} on the tail, whose parameters are summed up in Tab. \ref{tab:slopes}.
%    }
    }
    \label{fig:dead}
\end{figure}

\FloatBarrier
\begin{figure}
    \centering
    \includegraphics[width=\columnwidth]{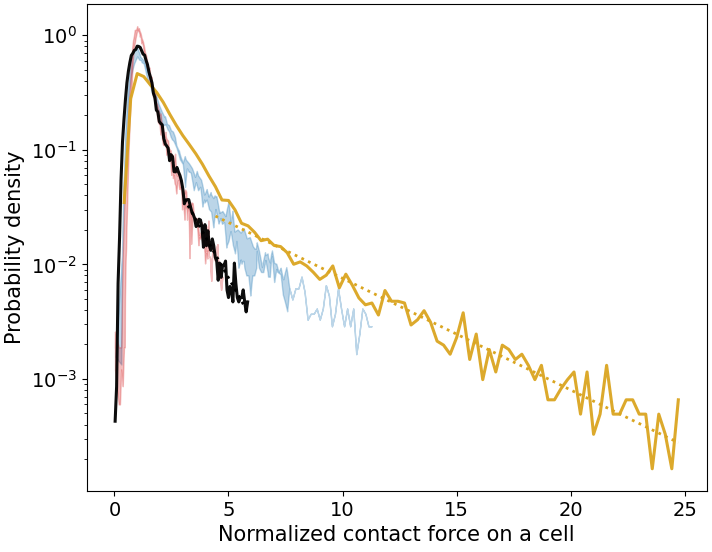}
    \caption{%
%    \textcolor{red}{
    Distribution of the normalized contact force of alive cells in the packed (black) and open (golden) spatial configurations without the beam (i.e. $100$\% alive cells). The broken lines in the tail and the red and blue shaded area, plotted for reference, follow the same convention of Fig.~\ref{fig:dead}.
%    }
    }
    \label{fig:alive100a}
\end{figure}

\section{Results}

\begin{table*}[t]
    \centering
    \begin{tabular}{|c|c|c|c|c|c|}
%    \hline
    \toprule
        \textbf{System} & $a$ \textbf{for alive cells} & $a$ \textbf{for dead cells} & $x_0$ \textbf{for alive cells} & $x_0$ \textbf{for dead cells} & \textbf{\% of alive cells} \\
        \hline \hline
        \multirow{4}{*}{Packed} & $ -0.215 \pm 0.007  $ & - & $-13.1 \pm 0.9$ & - & 100 \\
        \cline{2-6}
         & $ -0.42 \pm 0.01 $ & $ -0.63 \pm  0.07  $ & $-3.8 \pm 0.4$ & $-3.2 \pm 0.9$ & 78.8  \\
        \cline{2-6}
          & $ -0.52 \pm 0.03 $ & $ -0.85  \pm 0.06  $ & $-2.42 \pm 0.4$ &
$-1.4 \pm 0.4$ & 60 \\
        \cline{2-6}
         & $ -0.81 \pm 0.02 $ & $ -0.88  \pm 0.04$ & $-0.07 \pm 0.01$ & $-1.2 \pm 0.3$ & 38.7 \\
        \hline
        \hline
        \multirow{4}{*}{Open} & $ -0.92 \pm 0.02 $ & - & $-0.3 \pm 0.1$ &
- & 100 \\
        \cline{2-6}
         & $ -1.18 \pm 0.03 $ & $ -2.1 \pm 0.3 $ & $0.43 \pm 0.07$ & $0.0
\pm 0.3$ & 80.3 \\
        \cline{2-6}
         & $ -1.21 \pm 0.04 $ & $ -2.4 \pm 0.1 $ & $0.5 \pm 0.1$ & $0.5 \pm 0.1$ & 60.9 \\
        \cline{2-6}
          & $ -1.59 \pm 0.05 $ & $ -2.5 \pm 0.1 $ & $ 0.44 \pm 0.05$ & $0.44 \pm 0.09$ & 39.5 \\
        \hline \hline
    \end{tabular}
    \caption{Parameters for the exponential fits, $a$ and $x_0$ in Eq.~\ref{eq:fitExp}. The tails for the probability densities of contact forces were fitted by an exponential, Eq.~\ref{eq:fitExp}. We report here the parameters for each simulation both in the case of fixed volume (\textit{Packed}) and  open growth (\textit{Open}), and for four different dead-alive final ratios. There are no dead cells in the first row of each block since the beam was turned off (see the parameters in  Table~\ref{tab:params}).}
    \label{tab:slopes}
\end{table*}

The \textsc{CellSim3D}\cite{Madhikar2018} model for mechanobiological studies of growth, division and proliferation of cells has been successfully
validated against experimental data for the mitotic index and packing of epithelial cells~\cite{Madhikar2020}, and it has been used to study tissue growth under different conditions~\cite{Mazarei2022-vu,Mazarei2023-sq}. 
%\textcolor{red}{
To ensure reliability, the new features we introduced in \textsc{CellSim3D} were validated in two complementary ways:
%}
%
First, we compared the packing properties of cells to those obtained from experiments~\cite{jose} for elastic spheres and found very good agreement. We compared also the contact forces and their distributions, in particular with respect to the jamming transition. The distribution of forces displays an exponential tail, in agreement with experimental data from elastic shells~\cite{jose}. There are only very few force measurements in cellular systems~\cite{Steinwachs2016, Dolega2017}, but Trepat \textit{et al.}~\cite{Trepat2009} measured traction distributions using Madin-Darby canine kidney cells, finding that the peak of the distribution can be well approximated by a Gaussian, while the tail is an exponential, in excellent agreement with our model predictions. This will be discussed in detail in Section~\ref{sec:fit}.

A second validation was performed with respect to radiation treatment, with two simulation setups. The first setup was aimed at testing the radiation beam in a common pathology such as prostate tumor: direct comparisons were made with published experimental results for radiation (photon) treatment for the human prostate cancer PC-3 cell line~\cite{Kaighn1979}. The second setup was aimed at the investigation of the force distribution in the system for a broad range of final states, varying the percentage of alive cells and the size of the box. We found good agreement despite the varying experimental setups. The main parameters of the simulations described in the next sections are summed up in Table~\ref{tab:params}. We will discuss these validations in the next two sections.

\subsection{Contact force probability density}

Eight different simulations were executed using different sizes of the simulation box, and percentage of dead cells at the final step. The spatial configurations are labeled as \textit{open} in the case of an infinite box with open boundaries and \textit{packed} in the case of a rigid box filled at the level of more than 80\% at the final step of the simulation (confluence or close to confluence). 
%
%\textcolor{red}{
In the packed setup, the box side was chosen to be small enough to mimic confined growth conditions such as within dense tissue or restricted \textit{in vitro} assays. 
On the other side, the open setup, the box was defined so that the spheroidal cell aggregate developed without contacting the walls. This corresponds to a more physiologically relevant free-growth situation and also demonstrates that the model reproduces spheroidal growth  independent of boundary effects.
%}
%
The percentages of alive cells at the end of the simulation were, respectively, 100\% (beam absent), 80\%, 60\% and 40\%. We reserved, in each simulation, 6\% of the time steps for the system to relax after radiation delivery. Up to 
%
%\textcolor{red}{
$25 000$ cells were present in the final step of the packed setup and $45 000$ simulated in each case.
%}
 
%%
It has been shown that near the jamming transition the probability density of the inter-particle forces $P(f)$ decays nearly exponentially at large forces; the exponential tail is a signature of disorder and mechanical frustration. We observed that, as the packing density is increased, the tail of the distribution crosses over to a Gaussian indicating a more uniform force distribution. This is in line with previous experimental and simulation studies~\cite{jose}.

We extracted the probability density of the contact forces $P(f)$. The force $f$ is normalized by the mode of the distribution, such that the peak is at $f = 1$. The probability densities are renormalized after cutting off the last 1\% of the distribution, since the points in that range have a very low statistics and represent, in the case of alive cells, newborns which have anomalously high values caused by the conformation after the division algorithm. We fitted the peak of the distribution with a Gaussian function
\begin{equation}
g(A, \bar x,\sigma;\ x) = A e^{-\frac{(x - \bar x)^2}{2 \sigma ^2}}
\label{eq:fitGauss}
\end{equation}
and the tail with an exponential function
\begin{equation}
  f(x;\ a, x_0) = e^{a(x-x_0)}
\label{eq:fitExp}
\end{equation}
in order to to determine the stress transfer and the state of the system with respect to the point of the jamming transition. In Eqs.~\ref{eq:fitGauss}~and~\ref{eq:fitExp} above, $A$ and $a$ are the respective amplitudes, $\bar x$ the position of the peak and $\sigma$ the width of the Gaussian.

\subsubsection{Analysis of the fit \label{sec:fit}}

%\textcolor{red}{
The fitted parameters of the exponential model [Eq.~\ref{eq:fitExp}] together with the survival percentages are reported in Table~\ref{tab:slopes}. The normalized force probability densities of dead cells in the six simulations with the beam active are presented in Fig.~\ref{fig:dead}, while Fig.~\ref{fig:alive100a} compares the corresponding distributions of alive cells in the simulation without beam to those obtained with the beam active.
%}
%
Table~\ref{tab:slopes} shows interesting features: First, the steepness of the tails of the distributions is higher in the open systems than in the packed ones. This is expected, since the packed system is characterized by less space per particle, thus causing a more abundant population of stronger forces. The same steepness pattern is found when comparing dead and alive cells, in both spatial configurations. There are two interpretations:
\begin{itemize}
\item The shrinking of the dead cells, i.e., mimicking phagocytosis, was not complete. Therefore, the majority of the dead cells have some free space around them and are not completely packed with their neighbors, resulting in softer contact and lower peak forces.
\item Implementation of cell shrinking: As a cell gains internal pressure
when growing, it loses internal pressure when shrinking. Cells whose internal pressure is close to P$_{\text{dead}}$, the lower threshold, have a weaker pressure force $\vec{F}^P$, which points from the center of mass outward, and, in principle, prevents the cell from folding. Cells with lower pressure are more prone to becoming concave in some areas, decreasing
the contact area with the neighbors, and hence the total contact force on
the cell.
\end{itemize}

We also observed that the relative errors 
%
%\textcolor{red}{
of the parameters extracted from dead-cell data are higher than those extracted from
%}
%
alive cells. Moreover, the relative errors of the open space simulations are higher than those in the packed ones. These facts reflect the abundance of statistics at higher forces. In fact, the shorter the sample in the tail, the higher the probability for some bins of the histogram to contain only one or two points. The latter ones flatten the tail and augment the spread. This feature does not depend on the binning of the histogram and can be explained exactly as in the previous paragraph.

\begin{table}[t]
    \centering
    \begin{tabular}{|c|c|c|}
%    \hline
     \toprule
    \textbf{Parameter} & \textbf{Packed dead} & \textbf{Open dead} \\
    \hline \hline
    $\sigma$ & $0.26 \pm 0.02$ & $0.107 \pm 0.005$ \\
    \hline
    $A$ & $1.39 \pm 0.06$ & $3.12 \pm 0.07$ \\
    \hline
    $\bar x$ & $0.98 \pm 0.02$ & $1.011 \pm 0.004$ \\
    \hline \hline
    \end{tabular}
    \caption{
    Parameters for the Gaussian fit (Eq.~\ref{eq:fitGauss}) for dead cells both in the case of fixed volume (\textit{Packed}) and open growth (\textit{Open}), in the simulation with a final percentage of 40\% dead cells. $\sigma$, representing the width of the Gaussian, is related to the stress transfer and describes the status of our system with respect to the jamming transition. $\bar x$ is forced to be $\approx 1$ since the distributions are normalized to have the peak at $f=1$. We report the set of parameters fitted only for one dose (0.4712\,Gy, determining a final dead to total cells ratio of 40\%) since variations in the fit for other simulations with different final ratios are negligible and well within the statistics. }
    \label{tab:gauss_params}
\end{table}

%% PARAGRAPH OK

The parameter values from fitting to a Gaussian, Eq.~\ref{eq:fitGauss}, are shown in Table~\ref{tab:gauss_params}. We notice that in the case of fixed volume (packed) spatial configuration, the peak is more than twice wider than in the case of free growth (open boundary) having $\sigma _{\text{packed}} = 0.26 \pm 0.02$ and $\sigma _{\text{open}} = 0.107 \pm 0.005$, respectively. This is easily explained by the fact that in a free growth environment only few cells at the very core feel strong forces from their neighbors, while cells at the border proliferate freely and tend to expand the system and reduce contacts with their neighbors rather than compress the system. This is not possible when the system has filled all the available volume, and higher forces are experienced throughout the system. This fact is also expressed in the statistical error of $\sigma$.

The exponential tails and the widths of the Gaussians provide insight on the state of the system with respect to the jamming transition. When the system has little to no volume available, cell mobility decreases dramatically, and the cells stick together forming a compact granular tissue. In this case, the force distribution has a long exponential tail with a very marked slope difference from the initial, Gaussian peak. This is evident in the force distribution of dead cells in the packed configuration in Fig.~\ref{fig:dead}. A similar result can be found in Fig.~7b of~\cite{Madhikar2020a}, where the authors showed the force distribution of a system of \textit{bidimensional} cells with high friction and no free volume available. On the other hand, when the system has room to expand at its borders and little free space in its core due to the deflation of dead cells, the exponential tail is still present but way steeper and almost contiguous to the Gaussian peak. This is the case of the dead cells in the open space configuration in Fig.~\ref{fig:dead}. The system is still dense in its core, but cell mobility is still considerable and clusters of cells can move. This is again in agreement with~\cite{Madhikar2020a}, where the authors show in their Fig.~7a the same system but with null friction. Even if the volume available is still zero, the absence of friction facilitates independent cell movement and keeps the system at the jamming threshold.

\subsubsection{Shape of the distributions}

%{\color{red}
Figure~\ref{fig:dead} illustrates that the peaks of the force probability distributions in the packed configuration are systematically lower than those in the open configuration. 
This difference is a direct consequence of normalization: since the packed systems exhibit longer tails, the probability mass is spread over a wider domain, lowering the height of the peak at the mode $\bar{f}$. 

Both configurations display exponential tails, but the open systems decay more steeply.  In the packed setup the variance is reduced, since the entire voxel of cells experiences essentially the same mechanical stress, leading to a more homogeneous distribution. 
In contrast, in the open spheroid, the pressure felt by cells varies strongly with position -- cells near the center are under far higher stress than those at the periphery -- resulting in a broader spread of forces and a correspondingly sharper peak around the most probable value.
This behavior is consistent with mechanical analogies to jammed granular media, where interparticle force distributions are well-known to exhibit exponential tails \cite{Majmudar2005}. 
%Here, the analogy is used to rationalize the \emph{shape} of the probability density functions under crowding and jamming, rather than as a direct biological validation.

In addition, Fig.~\ref{fig:alive100a} shows the distributions from simulations without beam exposure (black and gold curves) in contrast to the spread with different beam doses. 
In this case, both configurations, and especially the open one, exhibit a markedly wider range of forces. 
The absence of dead cells with lower internal pressure, which in the irradiated systems tend to soften and partially fold their structures, removes a mechanism that would otherwise dampen the forces. 
As a result, forces much higher than the peak value, in some cases exceeding it by up to a factor of 25, become apparent in the no-beam simulations.

The mechanical interpretation of these results is consistent with experimental measurements of stress and force transmission in 3D cell systems. Steinwachs \emph{et al.}~\cite{Steinwachs2016} used three-dimensional traction force microscopy to map stresses generated by single cells embedded in biopolymer networks, revealing highly anisotropic and intermittent force propagation. The results features exponential-like force distributions similar to those observed in our simulations.
At the tissue scale, Dolega \emph{et al.}~\cite{Dolega2017} quantified the compressive stresses inside multicellular spheroids and demonstrated that the pressure increases toward the core, reducing cell proliferation and viability.
These experimental findings align well with our simulated packed configurations, where the central regions sustain higher mechanical stress and show a greater fraction of dead or jammed cells.
%}

\begin{table}[t]
    \centering
    \begin{tabular}{|c|c|c|c|}
%    \hline
     \toprule
    \textbf{Label} & $\alpha$ [Gy]$^{-1}$ & $\beta$ [Gy]$^{-2}$ & \textbf{Data  from} \\
    \hline \hline
    $\alpha\beta$-1 & 0.064    & 0.0167 & Deweese \textit{et al.}~\cite{prost_ab1} \\
    \hline
    $\alpha\beta$-2 & 0.241   & 0.067 & Algan  \textit{et al.}~\cite{prost_ab2}  \\
    \hline
    $\alpha\beta$-3 & 0.487  & 0.055 & Leith  \textit{et al.}~\cite{prost_ab3} \\
    \hline \hline
    \end{tabular}
    \caption{ Survival parameters from other studies, and used in the current simulations. The parameters $\alpha$ and $\beta$ describe the survival probability of a cell hit by a beam of dose $D$ according to Eq.~\ref{eq:surv}.
    }
    \label{tab:abprost}
\end{table}

\subsubsection{Comparison with experimental data}

%% PARAGRAH OK

We now compare our results with the experimental data of Jose \textit{et al.}~\cite{jose}. They prepared thousands of micron-sized elastic shells suspended in a solvent, applied different loads to the samples, and extracted the force distribution. They found an initial peak and a long exponential tail. The most impressive similarities are found between our distributions in the packed configuration in Fig.~\ref{fig:dead}, and their curves at low fractions of volume occupied by grain. This is expected, since the system in our simulation does not fill completely the volume at its disposal, but it is still compressed in five out of six sides of the cubic box.

%% PARAGRAH OK

There are, however, two notable differences. The first is that in our work,
the gradient between the peak and the initial value, at $f\approx 0$, is high, ranging from two to three orders of magnitude. In their case, the difference is small and, in some curves, displays a plateau. Their abundance of low forces is explained by the fact that, in the small force domain, they compute the forces with a linear equation, starting from the deformation of the grain. They observe a plateau when the fraction of space occupied by the grains is lower than 0.7. The hard sphere packing theorem~\cite{spher} states that the maximum volume occupied by \textit{hard} spheres packed in space is $\frac{\pi}{3\sqrt{2}}\approx 0.74$, while the limit for \textit{random} packing is $0.637$~\cite{scott1969}.

%% PARAGRAH OK

Therefore, since the density of \textit{soft} grains in Ref.~\cite{jose} is even smaller than the limit density of hard grains, and it is comparable with the threshold of randomly packed hard spheres, multiple soft grains in their experiment must have no deformation at all, giving abundance of points at low force. On the other hand, in our case, cells proliferate from an initial core, and any cell at the border of the system has at least one neighbor with which it is in contact, thus yielding a force contribution close to the mean and far from zero.

%% PARAGRAH OK

The second difference to the distributions in Ref.~\cite{jose} is the fact that they measured forces up to 4 times the mean force, while we detected values that were 7 times higher than our peak value $\bar{f}$ (which is proximal to the mean and set to unity in Fig.~\ref{fig:dead}). This fact, however, is explained by 
how the forces act In a packed configuration, neighbor cells are constantly pushed one against the other, and a linear repulsion $\vec{F}^R$ arises. Since the force has no upper bound, the repulsion force can become arbitrarily large, and in close packing it can easily reach  high values.

%% PARAGRAH OK

They also observed a Gaussian behavior away from the jamming point by augmenting the load, and therefore the fraction of space occupied by the grains. Although our open boundary simulations also show a behavior similar to a Gaussian right after the peak, the setup is too different to be reliable for a comparison, since there is no load on our system, and the packing in the core of the system is only due to cell-cell adhesion.

%% PARAGRAH OK

Trepat \textit{et al.}~\cite{Trepat2009} measured traction distributions in
canine kidney cells. Although we cannot compare their results quantitatively -- we have force distributions -- qualitative features can be compared. The data and fits shown in Fig.~\ref{fig:dead} are in excellent agreement with the traction data of Trepat \textit{et al.}~\cite{Trepat2009} which also shows Gaussian peaks and exponential tails.

\begin{figure}
    \centering
    \includegraphics[width=0.5\textwidth,trim={1.5cm 0 2.cm 1cm},clip ]{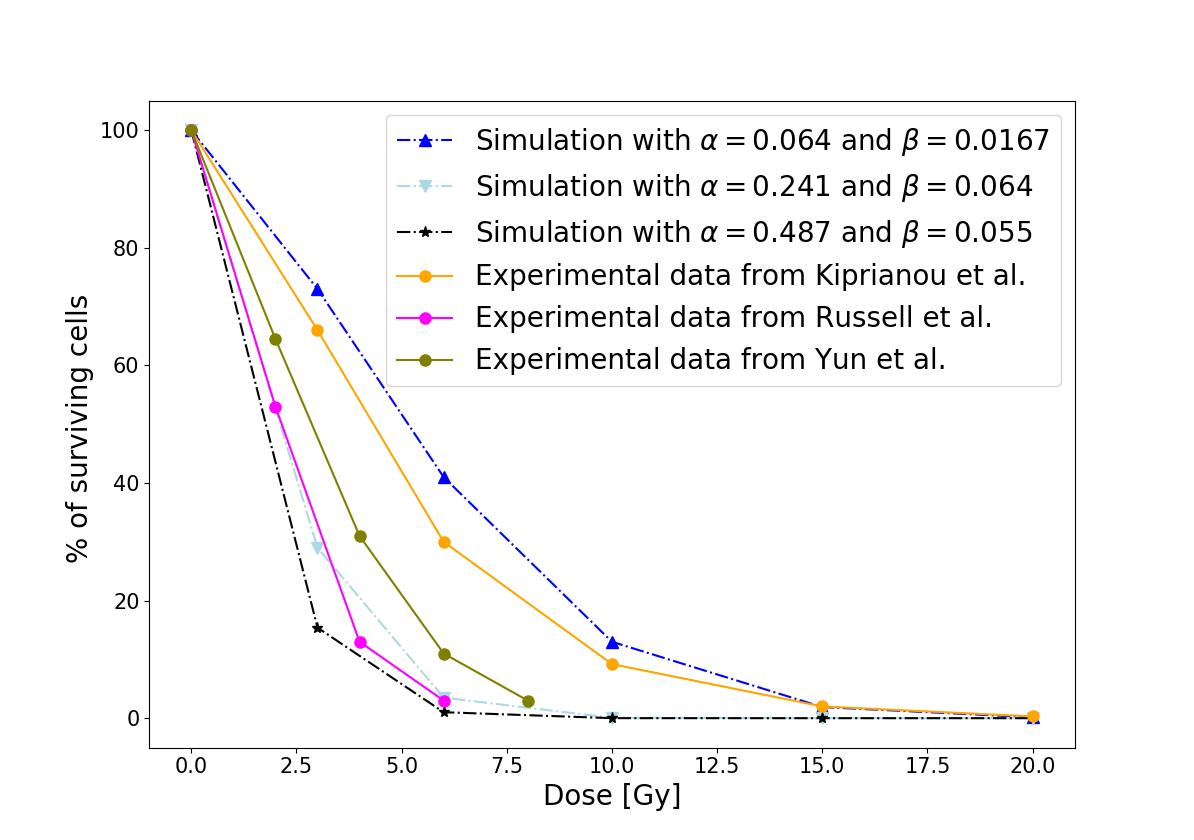}
    \caption{Percentage of surviving cells in our simulations (dashed lines), labeled with their  $\alpha$ and $\beta$ parameters (Table~\ref{tab:abprost}), and experimental data (solid lines) taken from Kiprianou  \textit{et al.}~\cite{prost_l1} (orange), Russell  \textit{et al.}~\cite{prost_l2} (magenta) and Yun  \textit{et al.}~\cite{prost_l3} (olive green).
  }
    \label{fig:prsurv}
\end{figure}

\subsection{Cell survival: the comparison with prostate cancer data}

%%%%%%%%%%%%%%%%%%%%%%%%%%%%%%%%%%%%%%%%%%%%%%%%%%%%%%%%%%%%%%%

%% PARAGRAH OK

Three data sets of survival parameters ($\alpha$, $\beta$) were extracted from the $\alpha/\beta$ ratios present in literature~\cite{prost_ab} along with three data sets~\cite{prost_l1, prost_l2, prost_l3} on the dead-to-alive ratios for the PC-3 cell line. In all cases radiation was delivered by a photon beam with no fractionation. We list the survival parameters
in Table~\ref{tab:abprost}.

%% PARAGRAH OK

The survival parameters $\alpha$ and $\beta$ do not depend on the dose $D$, but only on the nature of the beam and the structure of the tissue. However, these experimental values come from studies with different aims, namely the impact of the over-expression of a specific gene or a specific protein on the radiosensitivity of the tissue. Hence, these experimental data are influenced by external factors and the cell culture samples may also meet varying experimental conditions due to specific preparations, as listed below:
\begin{itemize}
\item In Kiprianou \textit{et al.}~\cite{prost_l1} the influence of an over-expression of a specific gene on radiosensitivity was carefully analyzed and compared with reference measurements performed on a pure PC-3 culture. Here, we used these unbiased reference data.
\item Russell \textit{et al.}~\cite{prost_l2} investigated the radiosensitivity enhancement caused by the expression of a specific protein. The study presents a control, in which the sample was mixed with a specific concentration of \textit{dimethyl sulfoxide} (DMSO) before irradiation. In this work, we used these unbiased data.
 \item In the case of Yun \textit{et al.}~\cite{prost_l3} the radiosensitivity enhancement was caused by the lack of a protein. Also in this case, we used the reference control, where the irradiated sample presents an endogenous level
of the protein.
 \end{itemize}
It is important to notice that, especially for improvements and accuracy, although we have chosen unbiased (or less biased) published data sets, differences in sample preparation, vector substance, and cell counting may complicate any quantitative comparison.

%% PARAGRAH OK

The simulation results and the experimental data extracted from references are plotted in Fig.~\ref{fig:prsurv}. As expected,  quantitative agreement is mixed: our results are mainly qualitative, since we collected parameters and values from previous studies having quite different experimental conditions. Quantitative agreement is restored at the tails of the distributions.

%% PARAGRAH OK

More specifically, we notice that the first set of parameters (blue curve in Fig.~\ref{fig:prsurv}) slightly overestimates the experimental set taken from Ref.~\cite{prost_l1} (orange curve) by a maximum $11$\% at $\text{D}=6$\,Gy, while the difference becomes lower than 8\% in all other cases. The disagreement further decreases toward the tail, as expected, since more cells are killed by the increased dose. The second set (cyan) and the experimental data taken from Ref.~\cite{prost_l2} (magenta), on the other hand, display a very good agreement, meaning that the measurements from Ref.~\cite{prost_l2} are well described by the parameters fitted in \cite{prost_ab2}. This is impressive, since the radiation is from a different source (respectively, X- and $\gamma$-ray) and a vector substance was added to the sample in the experimental setup, potentially slightly altering its radiosensitivity. The last set (black) describes a very sensitive
tissue and the simulation systematically underestimates the experimental data taken from Ref.~\cite{prost_l3} (green).

%% PARAGRAH OK

%
\begin{figure}[bt]
  \centering
%       \begin{subfigure}[b]{0.44\textwidth}
         \centering
         \includegraphics[width=0.23\textwidth]{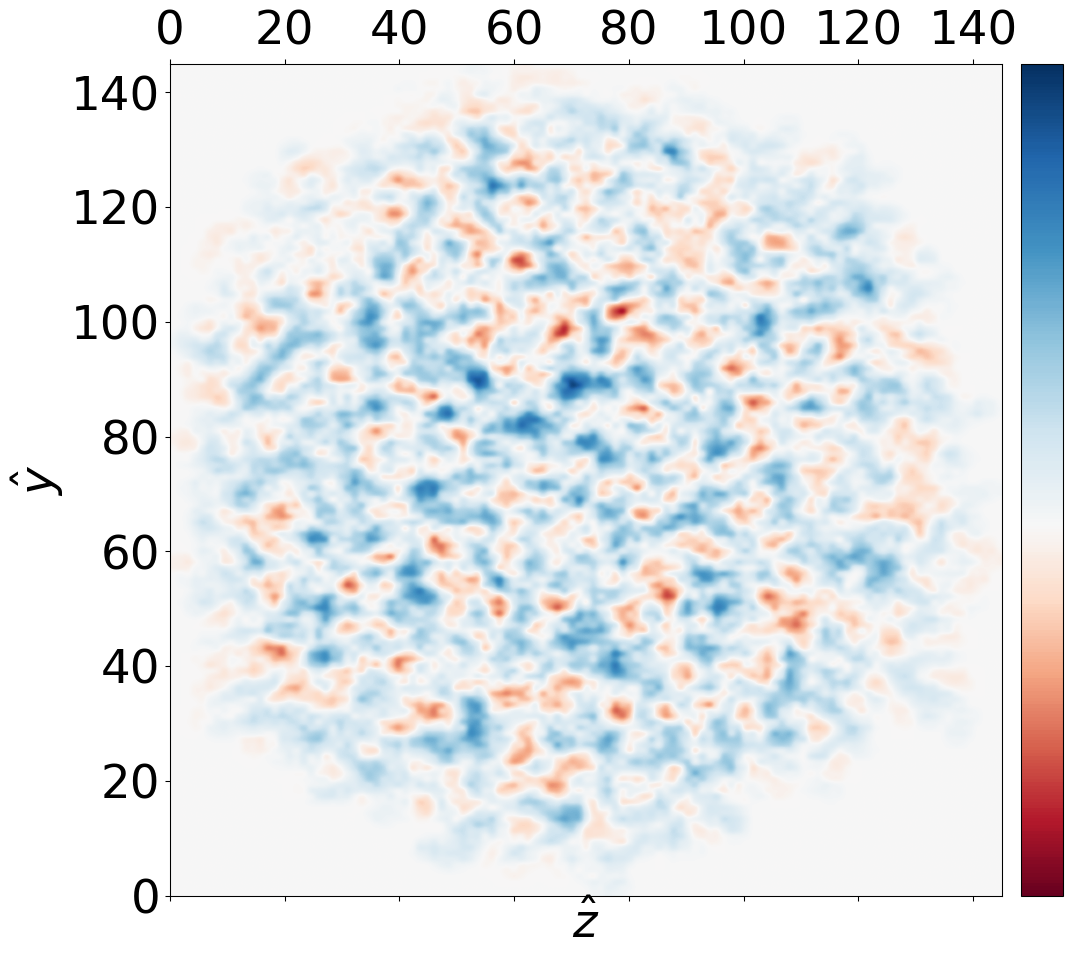}
%         \caption{Projection on $yz$ plane right after the beam.}
%         \label{fig:tc1}
%     \end{subfigure}
     \hfill
%       \begin{subfigure}[b]{0.44\textwidth}
%         \centering
         \includegraphics[width=0.23\textwidth]{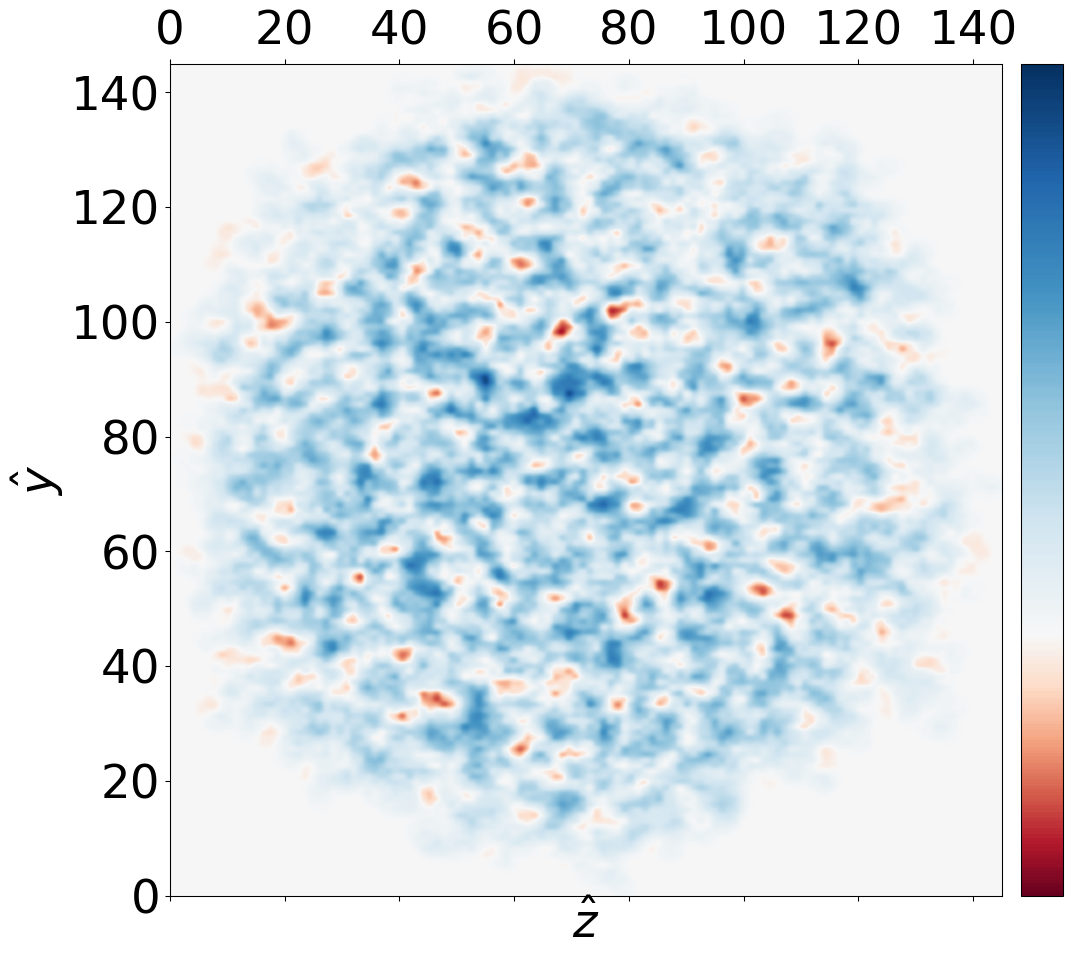}
%         \caption{Projection on $yz$ plane 2000 steps after the beam.}
%         \label{fig:tc2}
%     \end{subfigure}
     \hfill
%            \begin{subfigure}[b]{0.44\textwidth}
%         \centering
         \includegraphics[width=0.23\textwidth]{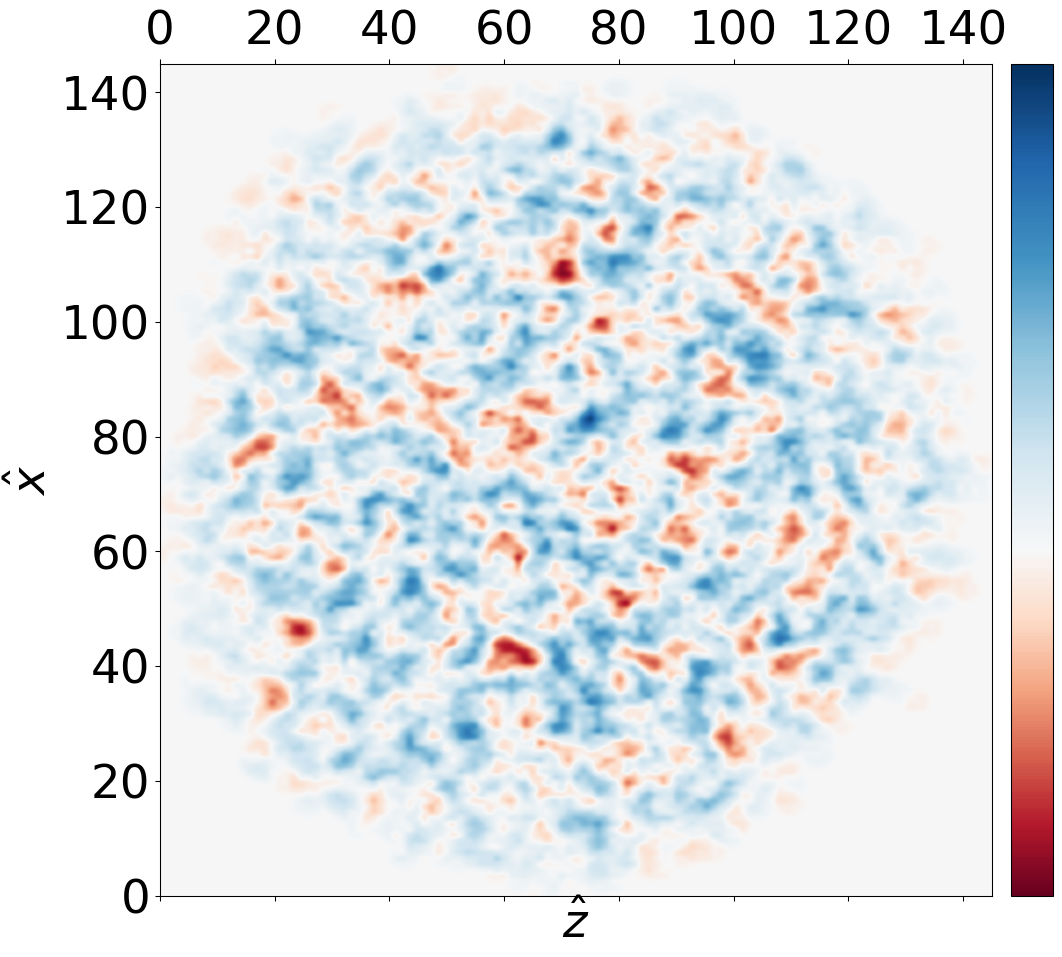}
%         \caption{Projection on $xz$ plane right after the beam.}
%         \label{fig:tc3}
%     \end{subfigure}
     \hfill
%            \begin{subfigure}[b]{0.44\textwidth}
%         \centering
         \includegraphics[width=0.23\textwidth]{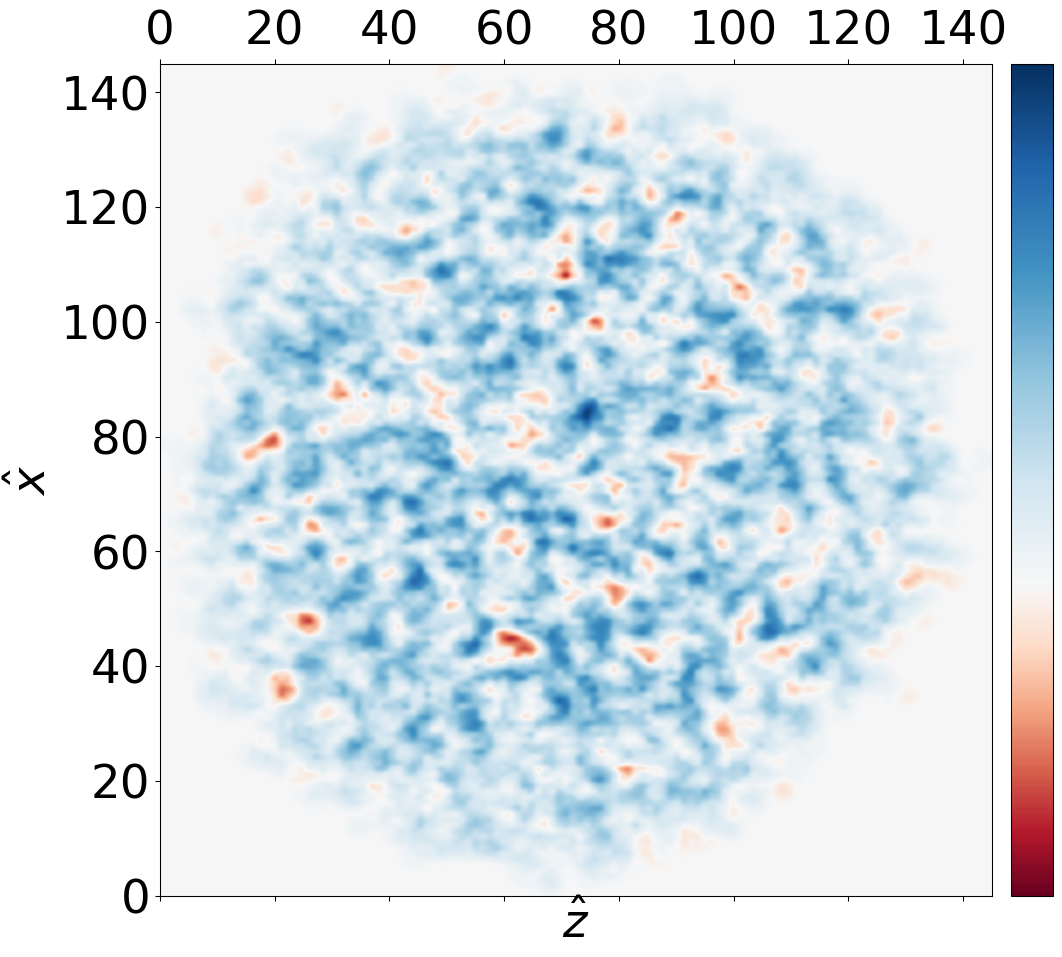}
%         \caption{Projection on $xz$ plane 2000 steps after the beam.}
%         \label{fig:tc4}
%     \end{subfigure}
%     \hfill
%     \begin{subfigure}[b]{0.44\textwidth}
%         \centering
         \includegraphics[width=0.23\textwidth]{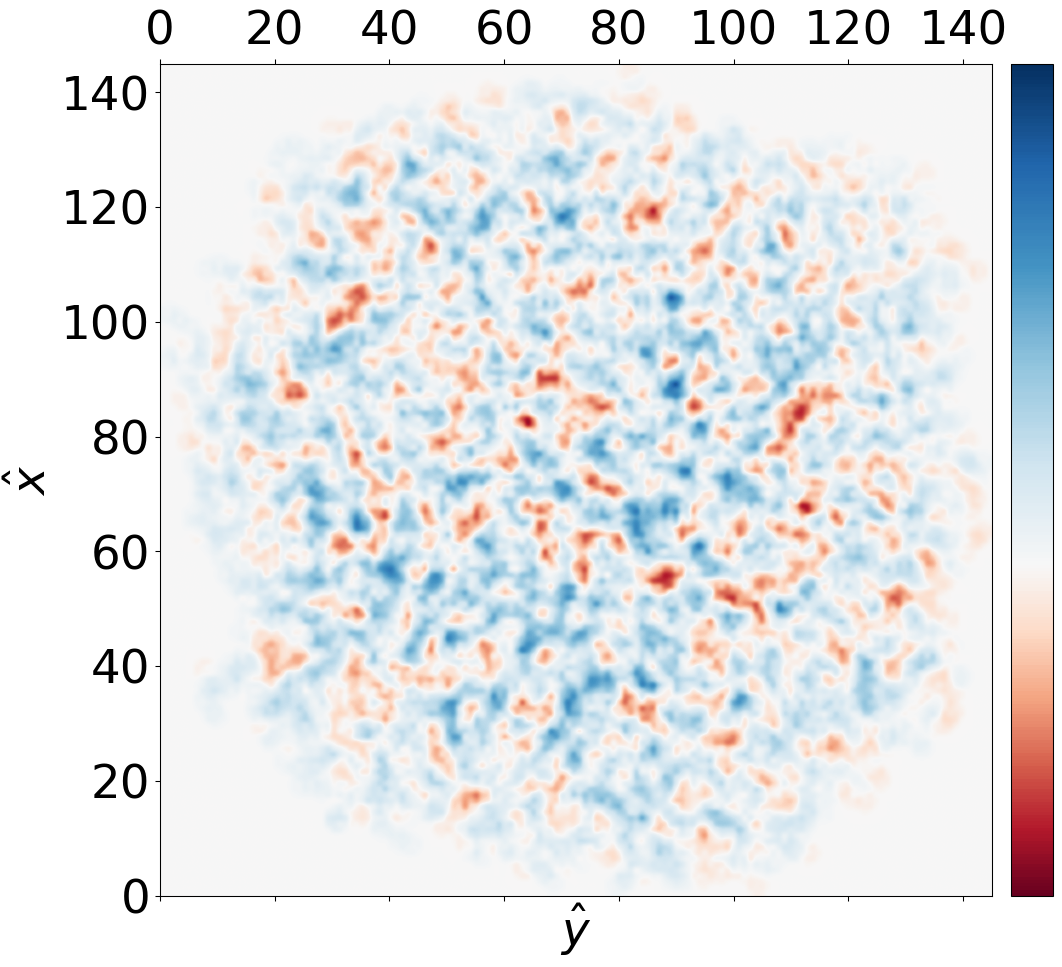}
%         \caption{Projection on $xy$ plane right after the beam.}
%         \label{fig:tc5}
%     \end{subfigure}
     \hfill
%     \begin{subfigure}[b]{0.44\textwidth}
%         \centering
         \includegraphics[width=0.23\textwidth]{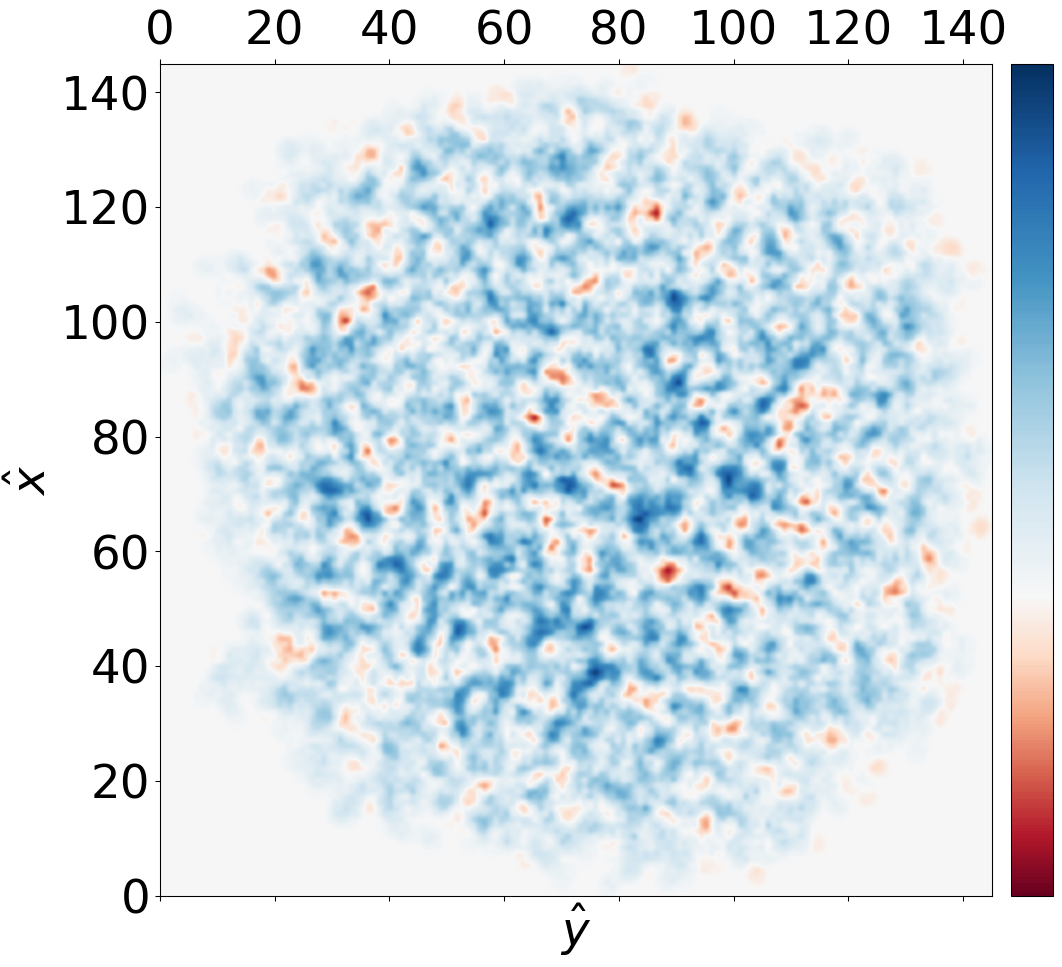}
%         \caption{Projection on $xy$ plane 2000 steps after the beam.}
%         \label{fig:tc6}
%     \end{subfigure}
    \hfill
%   \caption{Mass density projected along all the directions ($zy$-, $zx$- and $yx$-planes) both at the beam hit (left panels) and at the end (right panels) of the open boundary simulation with 60.9\% of alive cells at the final step. Blue: alive cells, and red: the dead cells. Proliferation tends to push the dead cells into clusters. The units in the axis are expressed in the natural units of the system (see Table~\ref{tab:params}).}
\caption{
%\textcolor{red}{
Mass density projected along all three orthogonal planes ($zy$-, $zx$-, and $yx$-planes) both at the beam hit (left panels), and at the end (right panels) of the open-boundary simulation with 60.9\% of alive cells at the final step. 
Blue represents alive cells and red the dead cells. 
Proliferation tends to push the dead cells into clusters as the system evolves. 
The units on the axes are expressed in the natural length scale of the model (see Table~\ref{tab:params}). 
A qualitatively similar spatial segregation of live and dead populations has been observed experimentally in 3D bioprinted tumor constructs combining cancer cells and cancer-associated fibroblasts, where confocal live/dead staining reveals clustered distributions of nonviable cells within the tissue~\cite{Baka2023}.
%}
}
    \label{fig:tumor_colormap}
 \end{figure}

In each simulation the final number of cells $n_0$  was between $ 43,000 $ and $ 45,000$, thus providing  sufficient statistics with $0.1\%$ standard deviation for all points. The experimental data were published without error bars, although the authors reported that all points represented mean values of at least three samples.

Figure~\ref{fig:tumor_colormap} shows  three spatial projections of the final two frames of the force distribution with open boundary configuration and with a $0.4712$ Gy dose (see Table~\ref{tab:params}). The mass densities of alive cells are  shown in blue and those of the dead cells in red. Clusters, or islands, of dead cells are clearly visible across the tumor after 2,000 steps of proliferation. The dead cells are pushed together by the division of the alive cells, which occurs even in the core of the system due to the free space releases by the dead cells.

\section{Conclusions}

We have proposed and analyzed a new tool for radiation oncology simulations by representing the tissue as a set of cells, rather than a continuum material, using the recently proposed \textsc{CellSim3D} mechanobiological cell model. We implemented the effects of radiation beam with a customizable dose, tissue sensitivity and fractionation as a stochastic Monte Carlo process. In a first set of runs, we measured the probability density $P(f)$ of the inter-cellular forces, finding an excellent agreement with experiments in many aspects, and some minor differences due to the cellular nature of the system. With these measurements, we were able to determine the state of the system with respect to the jamming transition, getting information on how far the cells can move inside the tumor. As a further proof of concept, we investigated the effect of radiation on prostate tumor model, finding good qualitative, but mixed quantitative agreement with experimental data. This is likely due to the varying experimental conditions present in literature. For further development, the software can be made to include spatial and temporal inhomogeneities for the dose and  tissue radiosensitivity as well as time dependence for, e.g., the dose rate. In addition, a more elaborated version of the survival probability can be used in place of Eq.~\eqref{eq:surv},  and cell migration can be used to simulate further dynamic effects of cancers, such as metastasis.  Such extensions would allow for more quantitative simulations of fractionated dose, inhomogeneous irradiation protocols, or even an inhomogeneous radiosensitivity of the tissue, and provide useful comparisons with experiments or even predictions for actual treatments.

\begin{acknowledgments}

We thank Francesco Cordoni for a critical reading of the manuscript.    
MK thanks the Discovery and Canada Research Chairs Programs of the Natural Sciences and Engineering Research Council of Canada (NSERC) and Foundation PS for financial support.

\end{acknowledgments}

%\bibliography{refs}

%%%%%%%%%%%%%%%%%%%%%%

%apsrev4-2.bst 2019-01-14 (MD) hand-edited version of apsrev4-1.bst
%Control: key (0)
%Control: author (8) initials jnrlst
%Control: editor formatted (1) identically to author
%Control: production of article title (0) allowed
%Control: page (0) single
%Control: year (1) truncated
%Control: production of eprint (0) enabled
%

\end{document}